\documentclass[preprintnumbers,amsmath,amssymb]{revtex4}
\usepackage{graphicx}
\usepackage{dcolumn}
\usepackage{bm}
\begin{document}
\def\oppropto{\mathop{\propto}} 
\def\opsimeq{\mathop{\simeq}}
\def\opoverderline{\mathop{\overline}}
\def\operarrow{\mathop{\longrightarrow}}
\def\opsim{\mathop{\sim}} 

\title{ Random polymers and delocalization transitions}
\author{ C\'ecile Monthus and Thomas Garel , \\
Service de Physique Th\'{e}orique, CEA Saclay,
91191 Gif-sur-Yvette cedex, France \\ }

 \affiliation{ Conference Proceedings ``Inhomogeneous Random Systems",
I.H.P., Paris, France, January 2006}

\begin{abstract}

\bigskip

In these proceedings, we first summarize
some general properties of phase transitions in
the presence of quenched disorder, with emphasis on the following
points: the need to distinguish typical and averaged correlations,
the possible existence of two correlation length exponents $\nu$,
the general bound $\nu_{FS} \geq 2/d$,
the lack of self-averaging of thermodynamic observables at criticality,
the scaling properties of the distribution of 
pseudo-critical temperatures $T_c(i,L)$ over the ensemble of samples
 of size $L$.
We then review our recent works on the critical
properties of various delocalization transitions involving random
polymers, namely (i) the bidimensional wetting (ii) the
Poland-Scheraga model of DNA denaturation (iii) the depinning
transition of the selective interface model (iv) the freezing
transition of the directed polymer in a random medium.

\end{abstract}

\maketitle



\section{Introduction}

In these proceedings, we review our recent works on the critical
properties of various delocalization transitions involving random
polymers, namely (i) the bidimensional wetting (ii) the
Poland-Scheraga model of DNA denaturation (iii) the depinning
transition of the selective interface model (iv) the freezing
transition of the directed polymer in a random medium. 

The paper is organized as follows.
In Section \ref{general}, we summarize some general properties of
random critical points.
The other Sections are devoted to the various models :
the wetting and Poland-Scheraga models
in Section \ref{wettps}, the selective interface model
in Section \ref{interface} and finally the directed polymer model
in Section \ref{dirpol}.

The remainder of this Introduction contains a brief
presentation of the various models we will discuss.

  \subsection{ Wetting and Poland-Scheraga models}
  
Wetting transitions are in some sense the simplest phase transitions,
since they involve linear systems \cite{mfisher}. Let us consider  a
one-dimensional random walk (RW) of $2L$ steps, starting at $z(0)=0$,
with increments $z(\alpha+1)-z(\alpha)=\pm 1$. The random walk
is constrained to remain  in the upper half plane $z \geq 0$, but
gains an adsorption energy $\epsilon_{\alpha}$ if $z(\alpha)=0$. More
precisely, the model is defined by the partition function 
\begin{equation}
Z_{wetting}(2L) = \displaystyle \sum_{RW} 
\exp \left( \beta \displaystyle \sum_{1 \leq \alpha \leq N}
 \epsilon_{\alpha}\delta_{z_{2 \alpha},0}  \right) 
 \label{zwetting}
  \end{equation}
with inverse temperature $\beta =1/T$.

In the pure case $\epsilon_{\alpha}=\epsilon_0$, there exists a 
continuous phase transition between
a localized phase at low temperature, characterized by an extensive
number of contacts at $z=0$, and a delocalized phase at high
temperature. The critical temperature is simply the point where
$e^{\beta_c \epsilon_0}=2$, i.e. where the energy gain ${\epsilon_0}$ of a
contact exactly compensates for its entropy loss $T_c \ {\rm Ln}2$.
At $T_c$, the wall $z=0$ is exactly reflexive, being attractive (resp.
repulsive) for $T<T_c$ (resp. $T>T_c$).

The Poland-Scheraga (PS) model of DNA denaturation \cite{Pol_Scher} is
closely related to the wetting model. It describes 
the configuration of the two complementary chains
as a sequence of bound segments and open loops.
Each loop of length $l$ has a polymeric entropic weight ${\cal N} (l)
\sim \mu^l /l^c $, whereas each contact at position $\alpha$ has a
Boltzmann weight $e^{- \beta \epsilon_{\alpha}}$. We assume that the
two chains are bound at $\alpha=1$ and $\alpha=L$. The partial
partition function $Z_{PS}(\alpha)$ with bound ends then satisfies the
simple recursion relation 
\begin{equation}
Z_{PS}(\alpha)=  e^{-\beta \epsilon_{\alpha} }  
  \sum_{\alpha'=1}^{\alpha-1}   {\cal N}(\alpha-\alpha') Z_{PS}(\alpha')
\label{recursion}
\end{equation}

The wetting model (\ref{zwetting}) corresponds to a Poland-Scheraga
model with loop exponent $c=3/2$ (this exponent comes from the first
return distribution of a one-dimensional random walk).  For DNA
denaturation, the 
appropriate value of the loop exponent $c$ has been the source of some
debate. Gaussian loops in $d=3$ dimensions are
characterized by $c=d/2=3/2$. The role of self avoidance
within a loop was taken into account by Fisher \cite{Fisher}, and
yields the bigger value $c=d \nu_{SAW} \sim 1.76$, where $\nu_{SAW}$
is the SAW radius of gyration exponent in $d=3$. More recently, Kafri
et al. \cite{Ka_Mu_Pe1,Ka_Mu_Pe2} pointed out that the inclusion of
the self avoidance of the loop with the rest of the chain further
increased $c$ to a value $c>2$. In the pure case
$\epsilon_{\alpha}=\epsilon_0$, the transition between the low
temperature bound phase and the high temperature unbound phase is
discontinuous for $c>2$, in marked contrast to the wetting case. The
discontinuous character of the transition was in fact previously found
in Monte Carlo simulations of self avoiding walks \cite{Barbara1}. The
value $c \simeq 2.11$ was subsequently measured \cite{Carlon, Baiesi1,
Baiesi2}.  

For the wetting and the Poland-Scheraga models, the question is
how  disorder in the contact
energies $\epsilon_{\alpha}$ modifies the critical
properties of the pure phase transition.

\subsection{Selective interface model}

Heteropolymers containing both hydrophobic and hydrophilic components
are of particular interest in biology. In a polar solvent, these
heteropolymers 
prefer conformations where the hydrophilic components
are in contact with the polar solvent, whereas hydrophobic
components avoid contacts with the solvent.
The behavior of such heteropolymers in the presence of an interface 
separating two selective solvents, one favorable to the hydrophobic
components and the other to the hydrophilic components,
is less obvious, and has been much studied recently.
In the initial work of ref. \cite{garel},
the following model was proposed for a polymer carrying random charges
 $\{q_{\alpha}\}$,  at the selective interface between two solvents,
 such that monomers with positive charges prefer to be in the upper
fluid (${\rm sgn }z >0$). The partition function 
\begin{eqnarray}
\label{definition}
Z_{SI}(L) = \sum_{RW}
exp \left(\beta \sum_{\alpha} q_{\alpha} {\rm sgn}z_{\alpha} \right)
\end{eqnarray}
 is over all random walks $\{z_{\alpha}\}$ of $L$ steps, 
with increments $z_{{\alpha}+1}-z_{\alpha}=\pm 1$ and (bound-bound)
boundary conditions $z_1=0=z_L$. Furthermore, it is convenient to choose
$q_{2{\alpha}+1}=0$ for all $(\alpha)$, so that there is no frustration at 
zero temperature (each monomer $z_{2{\alpha}}$ being in its preferred
solvent). Even charges $q_{2{\alpha}}$ are
random and drawn from the Gaussian distribution with mean value $q_0$
 \begin{eqnarray}
\label{gauss}
P(q_{2{\alpha}})= \frac{1}{ \sqrt{2 \pi \Delta^2} } e^{-
\frac{(q_{2{\alpha}}-q_0)^2} {2 \Delta^2} }
\end{eqnarray}
The initial work \cite{garel} found that for $q_0=0$ the chain
is localized around the interface at any finite temperature, whereas
for $q_0>0$ there is a phase transition at a critical temperature $T_c \simeq
O(\Delta^2/q_0)$ separating 
a localized phase at low temperature from a delocalized phase
into the most favorable solvent at high temperature.
Here, in contrast with the wetting and Poland-Scheraga models, 
there is strictly speaking no corresponding `pure' phase transition,
since an homogeneous chain $q_{\alpha}=q_0$ will be always
delocalized in its preferred solvent. To obtain a phase transition with
a non-disordered chain, one has to consider a periodic structure
of charges of both signs, the simplest case being an alternate sequence
$(q_A>0,-q_B<0)$ (see Ref \cite{gdcanonique,Bol_Gia} and references
therein). 
Another important difference with the wetting and Poland-Scheraga models
is that here the disorder is felt by all monomers,
whereas in the wetting and Poland-Scheraga models, the loops ($z_{\alpha}> 0$)
do not feel the disorder.

 \subsection{ Directed polymers in random media :
 transition towards a disorder-dominated phase }

 The model of directed polymer in a $1+d$ random medium 
 is defined by the following partition function
 \begin{equation}
Z_{L}(\beta) = \displaystyle \sum_{RW} 
\exp \left( \beta \displaystyle \sum_{1 \leq \alpha \leq L} 
\epsilon(\alpha, \vec r(\alpha))  \right) 
\label{directed}
  \end{equation}
over $d-$dimensional random walks $\vec r(\alpha)$, where the
 independent 
random energies $\epsilon(\alpha, \vec r)$ define the random medium.
This model has attracted a lot of attention because it is directly related
to non-equilibrium properties of growth models 
\cite{Hal_Zha}.
Within the field of disordered systems, it is also very interesting on its own
because it represents a `baby-spin-glass' model
\cite{Hal_Zha,Der_Spo,Der,Mez,Fis_Hus_DP}.  At low
temperature, there exists a disorder dominated phase, where the 
order parameter is an `overlap' \cite{Der_Spo,Mez,Com}.
In finite dimensions, a scaling droplet theory was proposed
 \cite{Fis_Hus_DP},
in direct correspondence with the droplet
 theory of spin-glasses \cite{Fis_Hus_SG},
whereas in the mean-field version of the model on the Cayley,
a freezing transition very similar to the one occurring
in the Random Energy Model was found \cite{Der_Spo}.
The phase diagram as a function of space dimension $d$ is the
following \cite{Hal_Zha}. In dimension $d \leq 2$, there is no free phase,
i.e. any initial disorder drives the polymer into the strong disorder phase,
whereas for $d>2$, 
there exists a phase transition between
the low temperature disorder dominated phase
and a free phase at high temperature  \cite{Imb_Spe,Coo_Der},
where the free energy has its annealed value.
 This phase transition  has been studied exactly on a Cayley tree \cite{Der_Spo}. In finite dimensions, bounds on the critical temperature
$T_c$ have been derived \cite{Coo_Der,Der_Gol,Der_Eva} :
$T_0(d) \le T_c \le T_2(d)$.
The upper bound $T_2(d)$ corresponds to the temperature above which the ratio
$\overline{Z_L^2}/(\overline{Z_L})^2$ remains finite as $L \to
\infty$. The lower bound $T_0$ corresponds to the temperature below which
the annealed entropy becomes negative. 

For the directed polymer model, one is thus interested both in the 
 properties of the low temperature disorder dominated phase
in any dimension $d=1,2,...$ and in the critical
properties of the transition that exists for $d \geq 3$. 

 \section{ General properties of random critical points (a physicist's
point of view)}
\label{general}
  
In this Section, we summarize the general properties of
 random critical points that will be useful to analyse
 the various random polymer models considered in the other Sections. 
  
  \subsection{ Harris criterion to determine disorder relevance near
second order pure critical points}

The stability of pure critical points with respect
to weak bond disorder is governed by the Harris criterion \cite{harris} :
near a second order phase transition in dimension $d$,
the bond disorder is irrelevant 
if the correlation length exponent $\nu_P \equiv \nu_{pure} > 2/d$,
or equivalently, using the hyperscaling relation $f \sim 1/\xi^d$,
if the specific exponent $\alpha=2-d \nu_P$ is negative $\alpha<0$.
On the contrary if $\nu_{P} < 2/d$ or $\alpha>0$, disorder is relevant
and drives the system towards a random fixed point
characterized by new critical exponents.

A simple argument to understand Harris criterion is the following.
The pure system at a temperature $T \neq T_c$ is characterized by
a correlation length $\xi(T) \sim t^{-\nu_P}$, where $t= \vert
T_c^{pure}-T \vert$ represents the distance to criticality, and
$\nu_P$ the correlation length exponent. The pure system can be
divided into nearly independent subsamples of volume $V \sim \xi^d(T)
\sim t^{-d \nu_P}$. In the presence of an additional weak bond
disorder, the averaged bond value $(1/V) \sum_{i \in V} J_i$ seen in a
volume $V$ will present fluctuations of order $1/\sqrt{V}$. So the
fluctuations of critical temperatures among the volumes of size $V$
will be of order $\Delta T_c(V) \sim 1/\sqrt{V} \sim \xi^{-d/2}(T)
$. Disorder will be irrelevant if these fluctuations $\Delta T_c(V)
\sim t^{d \nu_P/2}$ becomes negligeable with respect to $t=\vert T_c^{pure}-T
\vert$ in the limit $t \to 0$ where the critical point is approached.

\subsection{Correlation functions in disordered systems}

\label{corredis}

In disordered systems, it is well known that the partition function
$Z$ has a very broad distribution, which becomes peaked, in the
thermodynamic limit, around the typical (typ) value $Z_{typ} \sim
e^{\overline{{\rm Ln}Z}}$, whereas the averaged (av) partition
function $Z_{av}=\overline{Z}$ is usually atypical and dominated by
 rare samples. 

Correlation functions are, from this point of view, very similar to
partition functions. It is especially clear in one dimensional spin
systems \cite{Der_Hil,Cri,luck}, where correlation functions can be
expressed as product of random numbers. More generally in any
disordered system, the averaged correlation is expected to differ from the typical
correlation. However in contrast with partition functions
where the averaged value $Z_{av}$ has usually no physical meaning,
both the typical and averaged correlations are actually important, depending on the
physical quantities one wants to study \cite{Der_Hil,Cri,danielrtfic}.
The non self-averaging properties of correlations 
have also been studied in higher dimensional systems,
such as the two-dimensional (2D) McCoy-Wu model \cite{mccoywu,danielrtfic}
(see below), in 2D random $q$-state Potts model
\cite{Lud} and in the 3D random field Ising model \cite{Par_Sou}. 

As a consequence, the exponential decay at large distance
of the typical and averaged correlations lead to define two distinct correlation lengths.
It turns out that close to a phase transition, these two correlation lengths 
may have different critical behaviors.
 The best understood example of the existence 
of two different correlation length exponents is
 the random transverse field Ising chain 
(this quantum 1D model is equivalent to
 the 2D classical Ising model with columnar disorder introduced
 by McCoy and Wu \cite{mccoywu}),
 which has been studied in great details by D. Fisher 
via a strong disorder renormalization approach
 \cite{danielrtfic} : the exponent $ \nu_{typ}=1$ governs the decay of
the typical correlation at large distance $r$
\begin{equation}
\overline{ \ln C(r) } \sim - \frac{ r }{ \xi_{typ}}
\label{xityp}
\end{equation}
 whereas $\nu_{av}=2$ governs the decay of the averaged correlation
 \begin{equation}
\ln ( \overline{  C(r)} ) \sim -  \frac{r}  { \xi_{av} }
\label{xiav}
\end{equation}
 Exactly at criticality,
the typical and averaged correlations  are also very different, since
the typical correlation decays as $C_{typ} (r) \sim e^{- w { \sqrt
r}}$,  where $w$ is a random variable of order 1, whereas the averaged
correlation is dominated by rare events and decays algebraically $
\overline{  C(r)}  \sim 1/ r^{ (3-\sqrt 5)/2}$. 

 \subsection{ General bound $\nu_{FS} \geq 2/d$ for random systems, and
the possible existence of two distinct exponents $\nu$ }
 
There exists a general bound for the finite-size correlation length exponent
 $\nu_{FS} \geq 2/d$ in disordered systems \cite{chayes}, 
which essentially means
 that a random critical point should itself be stable with
 respect to the addition of disorder,
 as in the Harris criterion argument given above.
 However, this general bound has to be understood with the subtleties 
explained in \cite{chayes}.
 In so-called `conventional' random critical points, 
there is a single correlation length exponent
 $\nu=\nu_{FS}$ and this single exponent is expected to satisfy the bound.
 However, there are also `unconventional' random critical points, where
 there are two different correlation length exponents. 
In this case, the typical correlation
 exponent $\nu_{typ}$ can be less than $2/d$,
 whereas the bound holds for the finite-size exponent $\nu_{FS} \geq 2/d$.
 For instance in the random transverse field Ising chain where
 there are two diverging correlation lengths (Eqs \ref{xityp} and \ref{xiav}),
 the typical correlation exponent $ \nu_{typ}=1$ is less than $2/d=2$,
 whereas the finite-size exponent $\nu_{FS}=\nu_{av}=2$ that has to satisfy the bond $\nu_{FS} \geq 2/d=2$
 actually saturates it.
 Another important example discussed in \cite{chayes,danielrtfic}
 is the case of a first order transition that remains first order in the presence of quenched disorder :
 this first order transition in dimension $d$ is associated to the typical exponent $\nu_{typ}=1/d$,
 which is less than $2/d$, whereas the finite-size exponent saturates the bound $\nu_{FS}=2/d$. 
 The interpretation given in Sec. VII A of Ref. \cite{danielrtfic}
 is the following : the exponent $\nu_{typ}=1/d$
is expected to describe the rounding of the transition in a 
typical sample, whereas $\nu_{FS}=2/d$ describes the rounding of the transition
of the distribution of samples.  
Other critical points with two different correlation length exponents are discussed in \cite{singh,paz1,fisher2nu,bolech,myers}.

  \subsection{ Lack of self-averaging at random critical points}

 In disordered systems, the densities of extensive thermodynamic observables are self-averaging off-criticality,
because  the finiteness of the correlation length $\xi$
allows to divide a large sample into independent large
sub-samples.
At criticality however, this 'subdivision' argument breaks down
because of the divergence of $\xi$ at $T_c$, and a  
 lack of self-averaging has been found at criticality
 whenever disorder is relevant
\cite{domany95,AH,domany}. More precisely, for a given observable $X$,
it is convenient to define its normalized width as
\begin{equation}
\label{defratiodomany}
R_X(T,L) \equiv \frac{ \overline { X_i^2(T,L)} - ( \overline{X_i(T,L)})^2
}{ ( \overline{X_i(T,L)})^2 } 
\end{equation}
To be more specific, in ferromagnets, the observable $X$
can be the magnetization $M$, the susceptibility $\chi$,
the singular parts of the energy or of the specific heat \cite{domany}
In terms of the correlation length  $\xi(T)$, the following behaviour of
$R_X(T,L)$ is expected \cite{AH,domany}  

(i) off criticality, when $L \gg \xi(T)$, 
the system can be divided into nearly independent sub-samples
and this leads to `Strong Self-Averaging' 
\begin{equation}
R_X(T,L) \sim \frac{1}{ L^d} \ \ \hbox{ off  criticality  for 
 $L \gg \xi(T)$ } 
\end{equation}

(ii) in the critical region, when $L \ll \xi(T)$, 
the system cannot be divided anymore into nearly independent sub-samples.
In particular at $T_c(\infty)$ where $\xi=\infty$,
one can have either `Weak Self-Averaging' 
\begin{equation}
\label{weaksa}
R_X(T_c(\infty),L) \sim \frac{1}{ L^{d-\frac{2}{\nu_P}}} 
 \ \ \hbox{ for  irrelevant
disorder ($\nu_{P} > 2/d$)  } 
\end{equation}
or `No Self-Averaging'
\begin{equation}
\label{nosa}
R_X(T_c(\infty),L) \sim Cst \ \ \hbox{ for   random  critical points  }
\end{equation}
To understand the origin of this lack of self-averaging, it is useful
to introduce the notion of sample-dependent pseudo-critical temperatures,
as we now explain.

  \subsection{ Distribution of pseudo-critical temperatures}

Important progresses have been made recently in the understanding of
finite size properties of random 
critical points \cite{domany95,AH,paz1,domany,AHW,paz2,chamati} .
To each disordered sample $(i)$ of size $L$, one should first associate
a pseudo-critical temperature $T_c(i,L)$  \cite{domany95,paz1,domany,paz2}. 
Various definitions can be used, but one expects that
the scaling properties do not depend on the details of the definition.
For instance, in spin systems, one may define the pseudo-critical temperature $T_c(i,L)$
as the temperature $T$ where the susceptibility of the sample $(i)$ is maximum.
For the case of `conventional ' random fixed points that
are characterized by a single correlation length
exponent $\nu_R$,
the disorder averaged pseudo-critical critical
temperature $T_c^{av}(L) \equiv \overline{T_c(i,L)}$
satisfies 
\begin{equation}
T_c^{av}(L)- T_c(\infty) \sim L^{-1/\nu_{R}}
\label{meantc}
\end{equation}
which generalizes the analogous relation for
pure systems 
\begin{equation}
\label{puretc}
T_c^{pure}(L) - T_c(\infty) \sim L^{-1/\nu_{P}}
\end{equation}
The physical meaning of these equations is simply that 
a sample of length $L$ can be considered at criticality
when the correlation length $\xi \sim (T_c-T)^{-\nu_R}$ reaches the size $L$
of the system.

The nature of the disordered critical point then depends on the
width $\Delta T_c(L)$ of the distribution of the pseudo-critical
temperatures $T_c(i,L)$
\begin{equation}
\Delta T_c(L) \equiv \sqrt{Var [T_c(i,L)]}
=\sqrt{\overline{T_c^2(i,L)}-\left(\overline{T_c(i,L)}\right)^2}
\end{equation}
When the disorder is irrelevant, the fluctuations
of these pseudo-critical temperatures obey
the scaling of a central limit theorem as in the Harris argument : 
\begin{equation}
\Delta T_c(L) \sim L^{-d/2} \ \ \hbox{ for  irrelevant  disorder }
\label{deltatcirrelevant}
\end{equation}
This behaviour was first believed to hold in general \cite{domany95,paz1}, 
but was later shown to be wrong in the case of random fixed points.
In this case, it was argued \cite{AH,domany} that
eq. (\ref{deltatcirrelevant}) should be replaced by
\begin{equation}
\Delta T_c(L) \sim L^{-1/\nu_{R}} \ \ \hbox{ for  random critical points }
\label{deltatcrelevant}
\end{equation}
i.e. the scaling is the same as the $L$-dependent shift of the averaged 
pseudo-critical temperature (Eq. \ref{meantc}).
The fact that these two temperature scales remain the same
is then an essential property of random fixed points
that leads to the lack of self-averaging at criticality.

Up to now, to our knowledge, 
the distribution of $T_c(i,L)$ or of another sample-dependent
 critical parameter has been studied for
 various disordered spin models 
\cite{paz1,domany,paz2,igloipottsq}, 
for elastic lines in random media \cite{bolech},
for Poland-Scheraga models
\cite{PS2005}, for the selective interface model \cite{interface2005},
and for the directed polymer in a random medium of dimension $1+3$ \cite{future}.

  \subsection{ Finite-size scaling in disordered systems }

In pure systems, the finite-size scaling theory
 relates the critical exponents of the phase transition in the thermodynamic limit
to finite-size effects that can be measured in numerical simulations \cite{cardyFSS}.
In short, this theory says that the only important variable is
the ratio between the size $L$ of the finite system and
the correlation length that diverges at the critical point $\xi(T) \sim \vert  T-T_c \vert^{-\nu}$.
So the data $X_L(T)$ for various sizes $L$ should be analysed
in terms of the appropriate rescaled variable $\tau = (T-T_c) L^{1/\nu}$
to obtain a master curve of the form $ L^y X_L(T) =\phi( \tau)$,
where $y$ is the exponent describing the decay the observable
$X$ exactly at criticality $X_L(T_c) \sim 1/L^y$.
Note that using the scale-dependent $T_c(L)$ instead of the thermodynamic $T_c=T_c(\infty)$
is completely equivalent, since it 
corresponds to a simple translation $\tilde \tau= (T-T_c(L)) L^{1/\nu}= \tau +a $, as a consequence of (\ref{puretc}).

In random systems, one has instead data $X^{(i)}_L(T)$ measured at temperature $T$
for various disordered samples $(i)$ of size $L$, and the question is :
what is the best way to analyse these data?
The usual procedure consists in averaging over the samples $(i)$ at fixed $(T,L)$
to apply the pure procedure to these disorder averaged quantities :
one tries to find a master curve $ L^y \overline{ X^{(i)}_L(T) }  =\phi( \tau)$
in terms of the variable $\tau = (T-T_c) L^{1/\nu}$.
However, this procedure leads to extremely large sample-to-sample fluctuations
in the critical region, as a consequence of the width of the
distribution of pseudo-critical temperatures : at a given temperature $T$, the samples
having their pseudo-critical temperature $T_c(i,L)>T$
are effectively in the low temperature phase, whereas the samples having  
$T_c(i,L)<T$
are effectively in the high temperature phase. This mixing of samples
in the critical regions makes it very difficult to obtain
clean results on critical exponents.

To avoid these difficulties,
the following alternative procedure has been proposed \cite{paz1,domany,paz2} :
the data for a given sample $(i)$ should be analysed in terms
of the rescaling variable $ \tau_{new}= (T-T_c(i,L)) L^{1/\nu}$ with respect to
its own pseudo-critical temperature $T_c(i,L)$. 
Since $T_c(i,L)=T_c^{av}(L)+u_i \Delta T_c(L)$, where $u_i$ is a random variable
of order one, and where the mean $T_c^{av}(L) $ and the variance
$\Delta T_c(L)$
follow the respective behaviours (\ref{meantc}) and (\ref{deltatcrelevant}), it is clear that the two procedures
are not equivalent, since $\tau_{new}=\tau+a+u_i$ is not a simple translation of the constant $a$
as in the pure case, because the random variable $u_i$ remains present for arbitrary $L$.
This new way of analysing the data allows to reduce very significantly
the sample-to-sample fluctuations, as shown for spin models \cite{paz1,domany,paz2} ,
for the non-equilibrium depinning transition of
 elastic lines in random media \cite{olaf}, and for disordered polymer models \cite{PS2005}.

   \section{Disordered Poland-Scheraga model with various loop exponent $c$  }
\label{wettps}

\subsection{  Pure critical properties and disorder relevance : role of exponent $c$ } 

In the pure case $\epsilon_{\alpha}=\epsilon_0$, the model is of course exactly solvable,
and the critical properties are determined by the value of the loop exponent $c$ :
for $c>2$, the transition is first order with exponent $\nu_P=1$,
whereas for $1<c<2$ the transition is second order with exponent $\nu_P=1/(c-1)$.
A simple argument to understand these properties
is that the loop distribution at $T_c$ has a power-law decay involving
the exponent $c$ that enters the definition of the model
\begin{equation}
\label{probaloop}
P^{pure}_{Tc} (l) \sim \frac{1}{l^c}
\end{equation}
For $c>2$, the averaged length $<l> = \int dl \ l P^{pure}_{Tc} (l) $
is finite, so that the number $n(T_c)$ of contacts with the substrate
is extensive ($n(T_c) \sim L$); the transition is 
therefore first order. For $1<c<2$, the averaged length $<l>$ diverges,
and the L\' evy sum of $n$ independent variables $l_i$ drawn from the
distribution (\ref{probaloop}) scales as $l_1+...+l_n \sim n^{1/(c-1)}$.
As a consequence at criticality, the number of contacts $n^{pure}_L(T_c)$
in a sample of length $L$ scales as
\begin{equation}
n_L^{pure}(T_c) \sim L^{c-1}
\label{contactpure}
\end{equation}
and the transition is second order.

The Harris criterion concerning the stability
of pure second order transitions with respect relies on  the sign of the specific heat exponent 
\begin{equation}
\alpha_P=2-\nu_P = \frac{2c-3}{c-1}
\end{equation}
An equivalent way to decide whether disorder is relevant
consists in a simple power-counting analysis of the disorder 
perturbation exactly at $T_c$ :
the pure finite-size contact density $<\delta_{z_i,0}>_{pure} \sim
n_L^{pure}(T_c)/l \sim  L^{c-2}$
of Eq. (\ref{contactpure}) yields that the perturbation
due to the presence of a small disorder 
in the contact energies $\epsilon_i=\epsilon_0+\delta \epsilon_i$
scales as 
\begin{equation}
\label{powercounting}
 \sum_{i=1}^L \delta \epsilon_i <\delta_{z_i,0}>_{pure} \sim L^{1/2} \times L^{c-2}
= L^{c-\frac{3}{2}}
\end{equation}
Disorder is thus irrelevant for $1<c<\frac{3}{2}$
and relevant for $\frac{3}{2}<c<2$. 
Poland-Scheraga models are thus particularly interesting
to study disorder effects on pure phase transitions, since the parameter $c$
allows to study, within a single model,
 the various cases of second order transition with respectively
marginal/relevant disorder according to the Harris criterion,
or first-order transition.
From this point of view, it is reminiscent of the 2D Potts model,
where the pure critical properties vary with the parameter $q$ :
the transition is second order for $q<4$, the Ising case $q=2$
corresponding to the marginal case of the Harris criterion, whereas
the transition becomes first order for $q>4$.

 The marginal case $c=\frac{3}{2}$ has been studied for a long time
\cite{FLNO,Der_Hak_Van,Bhat_Muk,Ka_La,Cu_Hwa,Ta_Cha,wetting2005} 
and is of special interest since it corresponds to two-dimensional wetting
as explained in the Introduction. On the analytical side, efforts have focused on the
small disorder limit : Ref \cite{FLNO} finds a marginally
irrelevant disorder where
 the quenched critical properties are the same as in the pure case,
 up to subleading logarithmic corrections. Other
studies have concluded that that the disorder is marginally relevant
\cite{Der_Hak_Van,Bhat_Muk,Ka_La}. On
the numerical side, the same debate on the disorder 
relevance took place. The numerical studies of Ref. \cite{FLNO}
and Ref. \cite{Cu_Hwa} have concluded that the
critical behaviour was indistinguishable from the pure transition. On
the other hand, the numerical study of \cite{Der_Hak_Van}
 pointed towards a negative specific heat exponent
($\alpha<0$), and finally Ref. \cite{Ta_Cha}
has been interpreted as an essential singularity in the specific heat,
that formally corresponds to an exponent $\alpha=-\infty $.

 As explained in the Introduction, the case $c >2$ where the pure transition
is first order is of interest for DNA denaturation. The effect of
disorder on this transition has been recently debated
\cite{Barbara2,adn2005,Gia_Ton,PS2005}.

\subsection{Definition of a sample-dependent pseudo-critical temperature}

\label{deftcil}

In the magnetic systems studied in \cite{domany95,domany}, the pseudo-critical
temperature $T_c(i,L)$ of the sample $i$ was identified to the maximum
of the susceptibility. In the PS model, one can not follow the same
path and we have tried two different definitions in \cite{PS2005}.
Here we present the simplest one based on the free-energy.

In the pure PS model with bound-bound boundary conditions, 
the behavior of the partition function as a function of temperature reads
 \begin{eqnarray}
Z_L^{pure}&&( T<T_c) \opsimeq_{L \gg 1/(T-T_c)^{\nu_P}}
  (T_c-T)^{\nu_P-1} 2^L e^{ (T_c-T)^{\nu_P} L } \nonumber \\
 Z_L^{pure}&&(T_c) \simeq \frac{ 2^L }{ L^{2-c} }  
\label{znpur}  \\
Z_L^{pure}&&(T>T_c) \opsimeq_{L \gg 1/(T-T_c)^{\nu_P} }  \nonumber
 \frac{ 2^L }{ (T-T_c)^2 L^{c} } 
\end{eqnarray}
with $\nu_P=1/(c-1)$.
A finite-size pseudo-critical temperature $T_c^{pure}(L)$
can then be defined as the temperature where the free-energy
reaches the extensive delocalized value $F_{deloc}=-T L \ln 2$, i.e.
$T_c^{pure(f)}(L)$ is the solution of the equation
 \begin{eqnarray}
F_L^{pure}(L,T)+T L \ln 2=0
\end{eqnarray}
This definition  introduces a logarithmic factor
\begin{equation}
\label{lnfre}
T_c^{pure(f)}(L) - T_c(\infty) \sim \left( \frac{\ln L}{L} \right)^{1/\nu_{P}}
\end{equation}
with respect to the purely algebraic factor usually expected (Eq. \ref{puretc}).
This logarithmic factor comes the finite-size free-energy
value exactly at criticality $F_L^{pure}(L,T_c)=-T_c L \ln 2 +(2-c) T_c \ln L$
(Eq. \ref{znpur}). In the disordered case, we may similarly define a
sample-dependent pseudo-critical temperature $T_c^{(f)}(i,L)$ as the
solution of the equation
 \begin{eqnarray}
F_L^{(i)}(L,T)+T L \ln 2=0
\end{eqnarray}
Logarithmic corrections are to be expected 
to appear in the shift (Eq. \ref{meantc}).
This definition of the pseudo-critical temperature thus uses the fact
that the free-energy density of the delocalized phase is exactly known.

 \subsection{  Distribution of pseudo-critical temperatures in Poland-Scheraga models }

In \cite{PS2005}, the distribution of pseudo-critical temperatures 
disordered Poland-Scheraga models with different 
loop exponents $c$, corresponding to
either 
(i) a pure second order transition with marginal disorder $c=3/2$ (wetting case)  ;
(ii) a pure second order transition with relevant disorder $c=1.75$
(iii) a pure first order transition $c>2$ (DNA denaturation)

 In there three cases ($c=1.5$, $1.75$ and $2.15$),
the distributions of pseudo-critical temperatures
were found to follow the scaling form
\begin{equation}
P_L(T_c(i,L)) \simeq  \frac{1}{ \Delta T_c(L)} \  g \left( x= \frac{
T_c(i,L) -T_c^{av}(L)}{ \Delta T_c(L) }  \right) 
\label{rescalinghistotc}
\end{equation}
where the scaling distribution $g(x)$ is simply Gaussian
\begin{equation}
g(x)=  \frac{1}{ \sqrt{2 \pi} } e^{- x^2/2}
\label{gaussian}
\end{equation}
Note however that this Gaussian distribution is not generic
but seems specific to these Poland-Scheraga models, since in 
the selective interface \cite{interface2005}, the corresponding scaling distribution
was found to be very asymmetric.
The rescaling (\ref{rescalinghistotc}) means that the important scalings
of the pseudo-critical temperatures distribution
are the behaviours of its average $T_c^{av}(L)$ 
and width $\Delta T_c(L)$  as $L$ varies.

 For the marginal case $c=3/2$ corresponding to two-dimensional wetting, 
both the width $\Delta T_c(L)$ and the shift
$[T_c(\infty)-T_c^{av}(L)]$  are found to decay as $L^{-\nu_{R}}$, where the exponent is
very close to the pure exponent ($\nu_{R} \sim 2=\nu_{pure}$) but disorder is 
nevertheless relevant since it 
leads to non self-averaging of the contact density at criticality \cite{PS2005}.
For relevant disorder
$c=1.75$, the width $\Delta T_c(L)$ and the shift
$[T_c(\infty)-T_c^{av}(L)]$ decay with the same new exponent
$L^{-1/\nu_{R}}$ (where $\nu_{R} \sim 2.7 > 2 > \nu_{pure}$) and
there is again no self-averaging at criticality. Finally for the value
$c=2.15$, of interest in the context of DNA denaturation,  the width $\Delta T_c(L) \sim L^{-1/2}$ 
dominates over the shift $[T_c(\infty)-T_c^{av}(L)] \sim L^{-1}$,
i.e. there are two correlation length exponents $\nu=2$ and $\tilde
\nu=1$. This is reminiscent of what happens at strong disorder
fixed points \cite{danielrtfic,revue}, where the typical and averaged correlation
exponents are different as explained in Eqs (\ref{xityp}, \ref{xiav}).

 \subsection{ Discussion on the nature of the transition for
 $c=2.15$ in the presence of disorder}

\label{twoexponents}

The distribution of pseudo-critical temperatures for $c=2.15$ 
shows that the transition is an unconventional random critical point
with two different correlation length exponents
$\nu=2$ and $\tilde{\nu}=1$.
This is in contrast with usual random critical points, 
arising from second order transitions with
relevant disorder, where the same exponent is expected to govern 
 the width and the shift (Eq. \ref{deltatcrelevant}),
but this is reminiscent of what happens at strong disorder
fixed points \cite{danielrtfic,revue}.
The question is now which correlation exponent appears
in a given observable. In the random transverse field Ising
chain where many exacts results are known for exponents and
scaling distribution functions \cite{danielrtfic},
it is well understood how the two exponents $\nu=2$ and
${\tilde \nu}=1$ govern respectively the averaged/typical correlations.
Here in the disordered PS model, the analog of the correlation
function is the loop distribution. 
To simplify the discussion, let us more specifically
consider the probability of an end-to-end loop of length $L$ in sample (i)
of length $L$, which is directly related to the partition function
$Z^{(i)}_L(T)$ of sample $(i)$
\begin{equation}
P^{(i)}_L(L,T)=\frac {2^L}{L^c}\  \frac {1}{Z^{(i)}_L(T)}
\end{equation}

Introducing for each sample $(i)$ the
difference between the free-energy density 
$F^{(i)}_L(T)/L=-T \ln Z^{(i)}(L,T)/L$
 and the delocalized value $f_{deloc}=-T \ln 2$ 
\begin{equation}
\label{deffi}
f^{(i)}(L,T) \equiv \frac{-T \ln Z^{(i)}(L,T)}{L} +T \ln 2
\end{equation}
 one gets
\begin{equation}
\label{logpl}
{\rm ln }P^{(i)}_L(L,T)=-c \ {\rm ln} L + L \beta f^{(i)}(L,T)
\end{equation}
The self-averaging property of the free energy means
 that $f^{(i)}_L(T)$ converge for large $L$ to a non-random value
$f(T)$ for any sample $(i)$ with probability one
\begin{equation}
 f^{(i)}_L(T)  \operarrow_{L \to \infty}  f(T)
\end{equation}
where $f(T)$ is the free-energy difference between 
the localized phase and the delocalized phase :
$f(T<T_c)<0$ and $f(T>T_c)=0$.
This translates immediately into the corresponding statement (\ref{logpl})
for the logarithm of end-to-end loop probability
\begin{equation}
\frac{ {\rm ln }P^{(i)}_L(L,T) }{L} \operarrow_{L \to \infty} \beta f(T)
\end{equation}
for any sample $(i)$ with probability one.
Since the typical correlation length $ \xi_{typ}(T)$ 
is defined as 
 the decay rate of the logarithm of the correlation,
we obtain here that it is simply given by the inverse of the free-energy 
$f(T)$
\begin{equation}
\label{xitypfree}
 \frac{1}{ \xi_{typ}(T)} \equiv - \lim_{L \to \infty} 
\left( \frac{ {\rm ln }P^{(i)}_L(L,T) }{L}  \right) = - \beta f(T) 
\end{equation}

The dominance of the variance 
$\Delta T_c(L)\sim L^{-1/2}$ over the shift $[T_c^{av}(L)-T_c(\infty)]\sim L^{-1}$ 
indicates that asymptotically for large $L$, half of the samples $(i,L)$
are still localized at $T_c(\infty)$, whereas the other half
is already delocalized. This suggests that the contact
density is finite at criticality, as we have numerically
found in \cite{adn2005,PS2005}.  Similarly, the Monte-Carlo study
of 3D Self-Avoiding Walks with random pairing energies \cite{Barbara2}
point towards a finite energy density at $T_c$.
The free-energy of the disordered Poland-Scheraga model
is thus expected to vanish 
linearly
\begin{equation}
\label{freec215}
 f(T)  \operarrow_{T \to T_c^-} (T_c-T)
\end{equation}
The typical correlation length then involves the exponent $ \nu_{typ}=1$ (\ref{xitypfree})
\begin{equation}
\label{resxityp}
\tilde \xi(T)  \operarrow_{T \to T_c^-}
(T_c-T)^{-  \nu_{typ} } \ \ {\rm with } \ \   \nu_{typ}=1
\end{equation}

Let us now consider the decay of the averaged end-to-end loop distribution
that defines an a priori different correlation length $\xi_{av}(T)$
\begin{equation}
\label{lnav}
\frac{ {\rm ln } \left( \overline {P^{(i)}_L(L,T) } \right) }{L}
\operarrow_{L \to \infty} - \frac{1}{ \xi_{av}(T)}
\end{equation}
This correlation length $\xi_{av}(T)$ determines the divergence of
high moments of the averaged loop distribution.
At a given temperature $T<T_c$, these moments will actually
be dominated by the rare samples 
of length $L$ which are already delocalized at $T$, i.e. the 
samples having $T_c(i,L)<T$. Since our numerical results indicate that
the distribution of the pseudo-critical temperature $T_c(i,L)$
is a Gaussian with mean and width given respectively
by  $[T_c^{av}(L)-T_c(\infty)]\sim L^{-1}$ and $\Delta T_c(L)\sim L^{-1/2}$,
we obtain that the fraction of delocalized samples
presents the following exponential decay in $L$
\begin{equation}
{\rm Prob }[T_c(i,L)<T] \sim  e^{- (T_c^{\infty}-T)^2 L }
\end{equation}
This measure of the rare delocalized samples will govern  
the decay of the averaged
loop distribution, and the correlation length defined in (\ref{lnav})
thus involves the exponent $\nu_{av}=2$
\begin{equation}
 \xi_{av}(T) \operarrow_{T \to T_c^-} 
(T_c-T)^{ - \nu_{av}} \ \ {\rm with } \ \  \nu_{av}=2
\end{equation}
in contrast with the typical correlation length (\ref{resxityp}).

To better understand the emergence of two different correlation lengths,
we have numerically measured the distribution over the samples $(i)$
of the free-energy $f^{(i)}(L,T)$ defined in Eq. (\ref{deffi}).
We obtain that for $T<T_c$ 
\begin{equation}
f^{(i)}_L(T) = f(T)+\frac{a_T}{L}+\frac{\sigma_T u_i}{\sqrt L}
\end{equation}
where $a_T$ is temperature dependent and $u_i$ is a Gaussian random
variable of zero mean and of variance $1$ 
\begin{equation}
G(u) = \frac{1}{\sqrt{2 \pi } } e^{- \frac{u^2}{2 }}
\end{equation}
The averaged end-to-end loop distribution then reads (\ref{logpl})
\begin{eqnarray}
\overline {P^{(i)}_L(L,T)}  = \frac{1}{L^c}
  \overline{ e^{ L \beta f^{(i)}(L,T) } }
=  \frac{1}{L^c} e^{ L \beta f(T) }
 \int_{-\infty}^{+\infty} du \ G(u) 
 e^{ {\sqrt L} \beta \sigma_T u }  =  \frac{1}{L^c} e^{ L \beta f(T)
 +L \frac{\beta^2 \sigma^2_T }{2} }
\end{eqnarray}
The difference between
the correlation length $\xi_{av}(T)$ (\ref{lnav})
and the typical correlation length $\xi_{typ}(T)$ (\ref{xitypfree})
is due to the variance $\sigma^2_T$
\begin{equation}
\frac{1}{ \xi_{av}(T)} = \frac{1}{ \xi_{typ} (T)} - \frac{\beta^2 \sigma_T^2}{2} 
\end{equation}
In particular, to obtain the scaling $\frac{1}{ \xi_{av}(T)} \sim (T_c-T)^2$
different from $\frac{1}{ \xi_{typ}(T)} \sim (T_c-T)$, the variance term
in $\sigma_T^2$ has to cancel exactly the leading order
 in $(T_c-T)$ on the left hand-side.

So the picture that emerges of the present analysis is
very reminiscent of what happens at strong disorder fixed points
\cite{danielrtfic,revue} : the exponents $ \nu_{typ}=1$
and $\nu_{av}=2$ govern respectively the decay of
 typical/averaged loop distribution.
Our conclusion is thus that the exponent $\nu_{typ}=1$
governs the free-energy (\ref{freec215}) that corresponds to a Lyapunov 
exponent, i.e. it describes the critical behavior of
any typical sample,
whereas the exponent $\nu_{av}=2=2/d$ is the finite-size scaling exponent
of Chayes {\it et al} \cite{chayes} and is related to the variance of
the distribution of pseudo-critical temperatures.
The numerical results concerning the
contact density (Figure 8 of  \cite{adn2005}) may be now interpreted
as follows : for each sample, the critical region has a width
of order $1/L$, whereas the contact density averaged over the samples
decay on a much wider scale $1/\sqrt{L}$ that represents the 
sample-to-sample fluctuations of the pseudo-critical temperatures $T_c(i,L)$.

\section{Selective interface model}
\label{interface}  
  
  As explained in the introduction, the selective interface 
  model (\ref{definition}) is expected to undergo a
phase transition at a critical temperature $T_c \simeq
O(\Delta^2/q_0)$ between a localized phase and a delocalized phase in
the upper fluid. A real space renormalization group study, based on
rare events, was proposed in \cite{CM2000}. Mathematicians have also
been interested in this model. The localization at all temperatures
for the symmetric case was proven in ref. \cite{sinai,albeverio}. In
the asymmetric case, the existence of a transition line in
temperature vs asymmetry plane was proven in
ref. \cite{bolthausen,biskup}. More recent work can be found in
\cite{GT1,GT2,GT3,Gia_Ton}. 
In the following, we describe the results on the distribution
of pseudo-critical temperatures \cite{interface2005}.

\subsection{Definition of a sample dependent $T_c(i,L)$}

As for the wetting and Poland-Scheraga models, the free energy 
of the delocalized phase is known. If one forgets the boundary conditions at $(1,\alpha)$, 
the partition
function characterizing the delocalized phase in the $(+)$ solvent
($q_0>0$) would simply be for each sequence of charges
 \begin{eqnarray}
Z^{deloc}(\alpha) = 2^{\alpha-1} e^{\beta V(\alpha) }
\ \ \  { \rm with} \ \ \ 
 V(\alpha) \equiv  \sum_{\alpha'=2}^{\alpha} q_i
\label{zdeloc}
\end{eqnarray}
For each sample $(i)$ of length $L$, we may thus define a pseudo-critical
temperature $T_c(i,L)$ as the temperature where the free energy
$F^{(i)}(L,T) \equiv -T \ln Z^{(i)}(L,T)$
 reaches the delocalized value
 $F_{deloc}^{(i)}(L,T)=-T (L-1) \ln 2 -V(L)$ (Eq. \ref{zdeloc}),
i.e. $T_c(i,L)$ is the solution of the equation
 \begin{eqnarray}
\label{fre}
F^{(i)}(L,T)+T (L-1) \ln 2+V(L) =0
\end{eqnarray}

As explained in the previous section on Poland-Scheraga models,
this definition of pseudo-critical critical temperatures, together
with bound-bound boundary conditions, introduces logarithmic
correction in the convergence towards
$T_c(\infty)$. Eq. (\ref{meantc}) is accordingly replaced by 
\begin{equation}
T_c^{av}(L)- T_c(\infty) \sim \left( \frac{\ln L}{ L} \right)^{1/\nu_{R}}
\label{meantc2}
\end{equation}

\subsection{Scaling form of the probability distribution}

Our data for the distribution of pseudo-critical temperatures 
\cite{interface2005} follow the scaling form
\begin{equation}
P_L(T_c(i,L)) \simeq  \frac{1}{ \Delta T_c(L)} \  g \left( x= \frac{
T_c(i,L) -T_c^{av}(L)}{ \Delta T_c(L) }  \right) 
\label{rescalinghistotc2}
\end{equation}
where the scaling distribution $g(x)$ (normalized with $<x>=0$ and
$<x^2>=1$) is now very asymmetric, in marked contrast with 
the Gaussian form (\ref{gaussian})
obtained for the wetting and Poland-Scheraga models.

\subsection{Scaling properties of the shift and of the width}

We now discuss the numerical results obtained
 for the width $\Delta T_c(L)$ and for the
average $T_c^{av}(L)$ of the distribution (\ref{rescalinghistotc2}), as
$L$ varies. The width $\Delta T_c(L)$ follows the power law
\begin{eqnarray}
\label{reswidth}
\Delta T_c(L)  \sim   \left(\frac{1}{
L}\right)^{0.26} 
\end{eqnarray}
and the average $T_c^{av}(L)$ can be fitted
with the generalized form of eq. (\ref{meantc2})
\begin{eqnarray}
\label{resav}
T_c(\infty)-T_c^{av }(L)  \sim \left(\frac{\ln (L)}{L} \right)^{0.26}
\end{eqnarray}
The value $T_c(\infty) \simeq 0.838 \frac{\Delta^2}{q_0}$ is in
agreement with the numerical estimate of ref \cite{GT3}. The result
for the exponent in eqs (\ref{reswidth}) and (\ref{resav}) indicates
that the transition can be described as a random critical point with
a single correlation exponent 
\begin{equation}
\frac{1}{\nu_R} \simeq 0.26
\end{equation}
A similar value has been observed in numerical simulations by the
authors of ref. \cite{GT3} (private communication ).

Our data rule out the possibility of an infinite order transition
based on rare negatively charged sequences
that would lead to a smaller value of the critical
temperature $T_c^{rare}(\infty) = (2/3)
 \frac{\Delta^2}{q_0}$ \cite{CM2000}. The present results suggest that
the excursions in the unfavorable fluid that are important for
the transition, are of finite length.

 \section{ Freezing transition of the directed polymer in a random
medium} 
\label{dirpol}

We first summarize some properties of the low temperature phase, before
turning our attention towards the critical properties.

\subsection{ Statistics of excitations above the ground state}

The droplet theory for directed polymers \cite{Fis_Hus_DP},
is similar to the droplet theory of spin-glasses \cite{Fis_Hus_SG}.
It is a scaling theory that can be summarized as follows.
At very low temperature $ T \to 0$, all observables are governed by
the statistics of low energy excitations above the ground state.
An excitation of large length $l$ costs a random energy
\begin{eqnarray}
 \Delta E(l) \sim l^{\theta} u
\label{ground}
\end{eqnarray}
where $u$ is a positive random variable distributed with some law $Q_0 (u)$
having  some finite density at the origin  $Q_0 (u=0) >0$.
The exponent $\theta$ is the exponent governing 
the fluctuation of the energy of the ground state
is exactly known in one-dimension
$\theta(d=1)=1/3$ \cite{Hus_Hen_Fis,Kar,Joh,Pra_Spo}
and for the mean-field version on the Cayley tree
 $\theta(d=\infty)=0$ \cite{Der_Spo}.
In finite dimensions $d=2,3,4,5,...$, the exponent $\theta(d)$ has
been numerically measured, and we only quote here the results of the
most precise study we are aware of \cite{Mar_etal} for dimensions
$d=2,3$ : $\theta(d=2)=0.244$ and $\theta(d=3)  = 0.186$.

From (\ref{ground}), the probability distribution of 
large excitations $ l \gg 1$ reads within the droplet theory
\begin{eqnarray}
dl \rho (E=0,l)  \sim \frac{ dl }{l} e^{- \beta \Delta E(l)} 
\sim  \frac{ dl }{l} e^{- \beta l^{\theta} u  }
\label{rhodroplet}
\end{eqnarray}
where the factor $dl/l$ comes from the notion of independent excitations
\cite{Fis_Hus_SG}. In particular, its average over the disorder
follows the power-law
\begin{eqnarray}
dl \overline{ \rho (E=0,l) }  
\sim \int_0^{+\infty} du Q_0(u)  \frac{ dl }{l} e^{- \beta l^{\theta} u  }
= T Q(0) \frac{ dl }{l^{1+\theta}}
\label{rhoav}
\end{eqnarray}
This prediction describes very well the numerical data in 
the regime $1 \ll l \ll L$ in dimensions $d=1,2,3$ \cite{DPexcita}.

Since correlation functions at large distance are directly
related to the probability of large excitations,
we already see that the low temperature phase
is very non-trivial from the point of view of correlations
lengths : the typical exponential decay (\ref{rhodroplet}) indicates
a finite typical correlation length $\xi_{typ}(T)$,
whereas the averaged power-law behavior (\ref{rhoav}) means
that the averaged correlation length $\xi_{av}(T)$
is actually infinite in the whole low temperature phase
\begin{eqnarray}
\xi_{av}(0<T \leq T_c) =\infty
\end{eqnarray}
In addition to the general discussion of Section \ref{corredis},
this shows once again why it is crucial to distinguish
between typical and averaged correlation functions
in disordered systems. Note that within the droplet theory of 
spin-glasses \cite{Fis_Hus_SG}, the correlation length $\xi_{av}(T)$
is also infinite in the whole low temperature phase
for the same reasons.

\subsection{ Low temperature phase governed by a zero-temperature fixed point}

According to the droplet theory, the whole low temperature phase $0<T<T_c$
is governed by a zero-temperature fixed point. 
However, many subtleties arise because the temperature
is actually `dangerously irrelevant'. 
The main conclusions of the droplet analysis \cite{Fis_Hus_DP}
can be summarized as follows.
The scaling (\ref{ground}) governs the free energy cost
of an excitation of length $l$, provided one introduces
a correlation length $\xi(T)$ to rescale the length $l$
\begin{eqnarray}
\Delta F (l ) = \left( \frac{l}{\xi(T) } \right)^{\theta} u
\label{deltaF}
\end{eqnarray}
Here as before, $u$ denotes
 a positive random variable distributed with some law $Q (u)$
having  some finite density at the origin  $Q (u=0) >0$.
Moreover, this droplet free energy is a near cancellation of energy and
entropy contributions that scale as \cite{Fis_Hus_DP,Fis_Hus_SG}
\begin{eqnarray}
\Delta E (l ) \sim  l^{1/2} w
\label{deltaE}
\end{eqnarray}
where $w$ is a random variable of order $O(1)$ and of zero mean.
The argument is that the energy and entropy are dominated by small
scale contributions 
of random sign \cite{Fis_Hus_SG,Fis_Hus_DP}, whereas the free energy
is optimized on the coarse-grained scale $\xi(T)$. These predictions
for the energy and entropy have been numerically checked in
\cite{Fis_Hus_DP,Wa_Ha_Sc}.

\subsection{ Logarithmic fluctuations of the free energy at
criticality }

Let us now consider what happens for $T=T_c$.
Forrest and Tang \cite{Fo_Ta} have conjectured
from their numerical results on a growth model in the KPZ universality class
and from the exact solution of another model \cite{Blo_Hil}
that the fluctuations of the height of the interface
were logarithmic at criticality.
For the directed polymer model, this translates into
a logarithmic behavior of the free energy fluctuations at $T_c$
\begin{eqnarray}
\Delta F (L,T_c)  \sim (\ln L)^{\sigma} v
\label{fcriti}
\end{eqnarray}
where $v$ is a positive random variable of order one 
distributed with some law $R(v)$, and
where the exponent was measured to be in $d=3$ \cite{Fo_Ta,Ki_Br_Mo}
\begin{eqnarray}
\sigma = \frac{1}{2}
\label{sigma}
\end{eqnarray}

Further theoretical arguments
 in favour of this logarithmic behavior can be 
found in \cite{Ta_Na_Fo,Do_Ko}.
The argument of \cite{Do_Ko} is that the power-law behavior
$  F (L,T_c)  \sim L^{\theta_c}$ is impossible at criticality
so that $\theta_c=0$. From the scaling
relation $\theta_c=2 \zeta_c-1$  between exponents \cite{Hus_Hen},
the roughness exponent $\zeta$ is expected
to be exactly $\zeta_c=1/2$ \cite{Do_Ko}, 
and a renormalization argument then leads to logarithmic
 fluctuations of the free energy \cite{Ta_Na_Fo}.

\subsection{ Location of the critical temperature  }

\label{tct2}

\subsubsection{ Exact bounds on $T_c$ derived by Derrida and coworkers} 

Let us first recall the physical meaning of
the exact bounds for the critical temperature derived
by  Derrida and coworkers \cite{Coo_Der,Der_Gol,Der_Eva}
\begin{eqnarray}
T_0(d) \le T_c \le T_2(d)
\label{tcbounds}
\end{eqnarray}
The upper bound $T_2(d)$ corresponds to the temperature above which the ratio
\begin{eqnarray}
{\cal R}_L(T) = \frac{ \overline{ Z_L^2 } }{ ( \overline{ Z_L })^2 }
\label{ratiodef}
\end{eqnarray}
 remains finite as $L \to \infty$. The lower bound $T_0$ corresponds to the temperature below which
the annealed entropy becomes negative.

In dimensions $d=1,2$, the upper bound is at infinity $T_2=\infty$,
whereas  for $d \geq 3$, the upper bound $T_2$ is finite.
The interpretation is as follows \cite{Der_Eva}. The ratio (\ref{ratiodef})
can be decomposed according to the probability $P_L(m)$ that two
independent usual random walks in dimension $d$ meet $m$ times before
time $L$ 
\begin{eqnarray}
{\cal R}_L(T) 
= \sum _{m=1}^L P(m) B^m
\end{eqnarray}
where the factor
\begin{eqnarray}
B(T) =\frac{ \overline{ e^{2 \beta \epsilon}  } }{ ( \overline{ e^{ \beta \epsilon} } )^2 }
\end{eqnarray}
can be explicitly computed for any distribution of the site disorder
variable $\epsilon$. In dimensions $d=1,2$, two random walks meet an
infinite number of times as $L \to \infty$, whereas for $d \geq 3$,
they meet a finite number $m$ of times as $L \to \infty$. The
distribution of $m$ decays exponentially 
\begin{eqnarray}
P(m) \sim (1-A) A^m
\end{eqnarray}
where $(1-A)$ is the finite probability of never meeting again.
$T_2$ is defined as the temperature where 
\begin{eqnarray}
A B(T_2)=1
\end{eqnarray}
For $T>T_2$, $B(T)<B(T_2)=1/A$, and the ratio ${\cal R}_L(T)$ has a finite limit
\begin{eqnarray}
{\cal R}_{\infty}(T>T_2) = \frac{1-A}{1-AB(T)} 
\end{eqnarray}
For $T<T_2$, ${\cal R}_L(T)$ is a geometric series of parameter $A B(T) >1$,
and it thus diverges exponentially in $L$
\begin{eqnarray}
{\cal R}_L(T<T_2) 
\sim (1-A) \sum _{m=1}^L (A B(T) )^m \sim (A B(T) )^L
\end{eqnarray}
Exactly at $T_2$, the ratio diverges but not exponentially
\begin{eqnarray}
{\cal R}_L(T_2) 
= (1-A) \sum _{m=1}^L 1 \sim L
\label{ratiot2}
\end{eqnarray}

\subsubsection{ Interpretation  in terms of the probability distribution of free energies }

Let us now interpret the above results of the ratio ${\cal R}_L(T)$ 
 in terms of the probability distribution $P_L(F)$ of the free energy
$F=- kT \ln Z_L$ over the samples of length $L$. By definition (\ref{ratiodef}),
one has
\begin{eqnarray}
{\cal R}_L(T) = \frac{  \int dF P_L(F) e^{- 2 \beta F_L}  }{ ( \int dF P_L(F) e^{ - \beta F_L}  )^2 }
\label{ratiofreepdf}
\end{eqnarray}

For $T>T_2(d)$, the ratio ${\cal R}_{\infty}(T)$ is finite : this means that the fluctuations
of the free energy over the samples
\begin{eqnarray}
\left[ \Delta F_L \right]^2_{samples} =  \int dF P_L(F) F^2 - \left( \int dF P_L(F) F \right)^2
\end{eqnarray}
remain of order  $O(1)$ in the limit $L \to \infty$.

On the other hand, for the directed polymer in the low temperature phase $T<T_c$, 
the fluctuations of free energies over the samples is expected to have the same scaling
as the fluctuations of free energies within the same sample when the
end-point varies 
\cite{Fis_Hus_DP} : the fluctuations of free energy over the samples
are thus governed by the droplet exponent $\theta$
\begin{eqnarray}
\left[ \Delta F_L \right]_{samples} (T<T_c) \sim \left[ \Delta F_L \right]_{droplet} (T<T_c) \sim L^{\theta}
\label{samples-droplets}
\end{eqnarray}
Let us now recall Zhang's argument \cite{Hal_Zha} that allows to determine
the exponent $\eta$ of the tail of the free energy distribution
\begin{eqnarray}
P_L(F \to -\infty) \sim e^{- \left( \frac{ \vert F \vert}{L^{\theta}} \right)^{\eta} } 
\label{taileta}
\end{eqnarray}
Moments of the partition function can be then evaluated by the saddle-point
method, with a saddle value $F^*$ lying in the negative tail (\ref{taileta})
\begin{eqnarray}
\overline{ Z_L^n} = \int dF P_L(F) e^{ - n \beta F_L}  \sim \int dF 
e^{- \left( \frac{ \vert F \vert}{L^{\theta}} \right)^{\eta} } e^{ - n \beta F_L} 
\sim e^{ c(n) L^{ \frac{ \theta \eta}{1-\eta}   } }
\label{saddle}
\end{eqnarray}
Since these moments of the partition function have to diverge exponentially in $L$,
the exponent $\eta$ of the tail (\ref{taileta}) reads in terms of the droplet exponent
\begin{eqnarray}
\eta=\frac{1}{1-\theta} 
\end{eqnarray}

\subsubsection{ Debate on the location of $T_c$ for the directed polymer
 in finite dimensions }

At $T_c$, the fluctuations of the free energy are expected to be logarithmic,
as discussed around Eq (\ref{fcriti})
\begin{eqnarray}
\Delta F \sim (\ln L)^{\sigma}  \ \ \ {\rm with} \ \ \ \sigma=\frac{1}{2}
\end{eqnarray}
From these logarithmic fluctuations, 
 it seems rather difficult to obtain an exponential divergence in $L$ of the ratio ${\cal R}_L(T_c)$
 (\ref{ratiofreepdf}) if the free-energy distribution
$P_{T_c}(F)$ decays more rapidly than
exponentially as $F \to -\infty$.
On the contrary, if $T_c=T_2$, it
 is very natural to obtain the divergence found 
for the ratio at $T_2$ (\ref{ratiot2})
\begin{eqnarray}
{\cal R}_L(T_2)  \sim L \sim e^{ \ln L }
\end{eqnarray}
Moreover, to obtain the linear divergence (\ref{ratiot2}),
 the saddle-point method
described above for the low temperature phase  (\ref{saddle}) gives
that the tail of the free energy distribution should be at criticality
\begin{eqnarray}
P_{T_c}(F \to -\infty) \sim e^{- \left( \frac{ \vert F \vert}{ (\ln L)^{\sigma}} \right)^{\eta_c} } \ \ \ {\rm with } \ \ \eta_c=\frac{1}{1-\sigma} 
\end{eqnarray}
The value $\sigma=1/2$ corresponds to the tail exponent $\eta_c=2$.

We have thus proposed in \cite{DPdroplet}
 that the critical temperature $T_c$ in finite dimension $d$
satisfying $\theta(d)>0$
coincides with the temperature $T_2(d)$.
Explicit expressions for $T_2(d)$ in terms of usual integrals appearing
in the theory of random walks can be found in \cite{Coo_Der,Der_Gol}
for site and bond disorder respectively.

However, other arguments are in favor of the strict inequality
$T_c<T_2$ in finite dimensions. In particular 
 a new upper bound $T^*$ based on the entropy of the random
walk was recently proposed in \cite{birkner}. 
Moreover, the strict inequality $T_c<T_2$
is satisfied for the directed polymer
on the Cayley tree that plays the role of a mean-field version of the model
( $T_c$ coincides with the lower bound $T_0$ (\ref{tcbounds}) below which
the annealed entropy becomes negative \cite{Coo_Der} ).
It is useful to discuss the behavior of the free-energy distribution
 on the Cayley tree to compare with the finite dimensionsal case.
Let us first consider the low-temperature phase $T<T_c$.
On the Cayley tree, the low-temperature phase is characterized by
$\theta=0$ and $\Delta F =O(1)$ 
whereas Zhang's argument above is consistent only if $\eta =1/(1-\theta)
>1$ to ensure the convergence in the presence of the exponential term
$e^{- \beta n F}$ (Eq \ref{saddle}). When $\theta=0$, the tail of
$P_L(F \to -\infty)$ is
also an exponential $e^{ a F/(\Delta F)}$ as in the Random Energy Model
\cite{Der_Spo} and one has to take into account the minimal
free energy that can be obtained for a finite size $L$.
From a physical point of view, the reason could be that
the configurations of two polymers in the same disordered sample
are very different. In finite dimensions, contacts and loops
alternate extensively, whereas on the tree, the loops simply
do not exist : the two polymers may only coincide over some distance
and then never meet again. Since the exponential tail found for the
Cayley tree actually corresponds to the universal Gumbel tail for the
minimum of independent variables, this shows that the non-exponential
tail found in finite dimensions for $T<T_c$ reflects the importance of
correlations between the free energies of paths due to the presence of
loops. Let us now discuss what happens in the high temperature phase $T>T_c$.
On the Cayley tree, the tail of the free-energy distribution
$P_L(F \to -\infty)$ is
known to be also an exponential $e^{ a F/(\Delta F)}$ \cite{Der_Spo},
and this is why the ratio (\ref{ratiofreepdf}) 
can diverge exponentially in the region $T_c<T<T_2$ even if $\Delta F=O(1)$.
In finite dimension $d$, the free-energy fluctuations are expected to
be of order $\Delta F=O(1)$ for $T>T_c$. The debate
between the two possibilities $T_c=T_2$ or $T_c<T_2$ in finite dimensions
thus depends of the tail the free-energy distribution
$P_L(F \to -\infty)$ for $T>T_c$. If the tail is exponential
 as on the Cayley tree,
then $T_c<T_2$, whereas if $P_L(F \to -\infty)$ decays more
 rapidly than exponentially, then $T_c=T_2$.
For instance, in the mean-field Sherrington-Kirkpatrick model of 
spin-glasses, one has $T_c=T_2$ and the distribution of the free-energy
fluctuations for $T>T_c$ is known to be Gaussian 
\cite{Aiz_Leb_Rue,Com_Nev}. 
We are presently studying the directed polymer in $d=3$ 
numerically \cite{future}. Various indicators based on
geometrical or thermodynamical behaviors points towards
a critical temperature $T_c$ in the vicinity of $T_2$,
and thus the possibility $T_c=T_2$ cannot be 
presently excluded within the numerical precision.
Concerning the free-energy distribution, we obtain numerically
that it is Gaussian in the high temperature phase for $T \gg T_2$,
but we are not aware of any argument discussing the form of this
 distribution in
the literature.
In conclusion, we have the feeling that the most clear way
to solve the debate on the value of $T_c$ in finite dimension
would be to obtain results on the probability distribution of the
free-energy in the whole high-temperature phase $T>T_c$.

\subsubsection{ Why $T_c$ is different from $T_2$ in other finite dimensional
disordered systems  }

The fact that the fluctuations of free energies over the samples have the same scaling
as the droplet excitations within one given sample (\ref{samples-droplets})
is very specific to the directed polymer model.
In other disordered models, such as spin-glasses for instance, the fluctuations
of free energies over the samples scale instead as \cite{We_Ai,Bou_Krz_Mar}
\begin{eqnarray}
\left[ \Delta F_L \right]_{samples}  \sim  L^{d/2}
\label{clt}
\end{eqnarray}
at any temperature.
This scaling simply reflects the Central-Limit fluctuations of the $L^d$ disorder variables
defining the sample. The directed polymer escapes from these normal fluctuations
because it is a one-dimensional path living in a $1+d$ disordered sample :
each configuration of the polymer only sees $L$ random variables
among the $L^{1+d}$ disorder variables that define the sample,
and the polymer can 'choose' the random variables it sees.
So for other disordered systems having fluctuations over the samples governed by (\ref{clt}),
the ratio ${\cal R}_L(T)$ will diverge exponentially as any temperature.
The temperature $T_2$ is thus infinite
\begin{eqnarray}
T_2=\infty
\end{eqnarray}
and has nothing to do with any critical temperature.
However, the droplet exponent $\theta$ is expected to govern the
correction to the extensive part of the mean value \cite{Bou_Krz_Mar}
\begin{eqnarray}
\overline{ F_L }  \sim  L^{d} f_0 + L^{\theta} f_1
\label{faveraged}
\end{eqnarray}
It can for instance be measured in the free energy difference
upon a change of boundary conditions that forces the introduction
of some domain wall in the sample \cite{Fis_Hus_SG}.

\subsection{ Description of the transition in terms of the loop
distribution between two polymers in the same sample }

For $T<T_c$, the number of contacts of two independent polymers
$x(i)$ and $y(i)$
in the same disordered sample 
\begin{eqnarray}
n_L(T) = \sum_{i=1}^L < \delta_{x(i),y(i)}>
\label{ncontact}
\end{eqnarray}
is extensive, and the density of contacts, also called the overlap,
is precisely the order parameter of the low temperature phase
\cite{Der_Spo,Mez,Com}
\begin{eqnarray}
q(T) = \lim_{L \to \infty} \left( \frac{  n_L(T) }{L} \right)
\label{defoverlap}
\end{eqnarray}

Note that on the Cayley tree where $\theta=0$, the distribution
of this overlap is made of two delta peaks at $q=0$ and at $q=1$ 
\cite{Der_Spo},
whereas in finite dimensions with $\theta>0$, the distribution
of this overlap is expected to be a single delta function
at $\overline{q(T)}$ \cite{Mez}.

One may thus analyse the configuration of two polymers in the same
sample in terms of contacts separated by loops.
For $T<T_c$, the distribution of large loops follows
a scaling form based on the free energy scaling of a droplet
of length $l$ (\ref{deltaF}) 
\begin{eqnarray}
dl P_{large}(l,T) =
 {\cal N}(T)  \frac{dl}{l}
 e^{- \beta \Delta F (l )  }
= {\cal N}(T)  \frac{dl}{l}
 e^{- \beta \left( \frac{l}{\xi(T) } \right)^{\theta} u  }
\label{loop}
\end{eqnarray}

At $T_c$, the logarithmic scaling (\ref{fcriti})
suggests that the loop distribution still exists
and follows the form 
\begin{eqnarray}
dl P_{T_c}(l) =  \frac{dl}{l}  e^{- \beta \Delta F(l)  }
=\frac{dl}{l}  e^{- \beta_c (\ln l)^{\sigma} v }
\label{looptc}
\end{eqnarray}

The normalization factor $ {\cal N}(T)$ of large loops
of the low temperature phase ( Eq. \ref{loop}) 
can be found by a matching procedure on scale $l \sim \xi(T)$
with the critical distribution (\ref{looptc})
as explained in details in \cite{DPdroplet}.
 For the value 
$\sigma=1/2$ measured in $d=3$ \cite{Fo_Ta,Ki_Br_Mo},
the resulting critical behavior for the free-energy and the overlap
are respectively
\begin{eqnarray}
 f(T)-f(T_c) &&  \sim \frac{1}{ \xi (T)  }  \sim 
e^{  -  \left( \frac{2}{K} \ln \frac{1}{T_c-T} \right)^{2} +... }  \\
q(T)  &&   \sim 
  e^{ -  \left( \frac{2}{K} \ln \frac{1}{T_c-T} \right)^{2}
+ 2 \ln \frac{1}{T_c-T}+... }  
\label{lasteq}
\end{eqnarray}
where $K$ is some constant.

In conclusion, the logarithmic behavior (\ref{fcriti}) with $\sigma=1/2$
is responsible for the unusual critical properties (\ref{lasteq}),
whereas the usual power-laws in pure phase transitions
correspond to the value $\sigma=1$.
Since the droplet scaling theory of the low temperature phase
was initially developed for spin-glasses \cite{Fis_Hus_SG},
this raises the question of the existence of an exponent
 $0<\sigma<1$ at the spin-glass transition. 
Some consequences of this possibility are discussed in \cite{DPdroplet}.
  
  \section{ Conclusion}

In these proceedings, we have first summarized
some general properties of phase transitions in
the presence of quenched disorder.
We have then reviewed our recent works on the critical
properties of various delocalization transitions involving random
polymers. In the wetting and Poland-Scheraga models, 
the delocalization transition already exists
in the pure case, and pure critical properties
depends on the value of the loop exponent.
The Harris criterion is a convenient measure of
the relevance of the disorder, and we have presented results
for second order with either marginal/relevant disorder,
as well as for first order with relevant disorder.
 In the selective interface model, we have explained why the scenario based
on rare large loops in the minority solvent seems now numerically ruled out.
 Further work is thus needed to elucidate the precise mechanism of 
the transition. Finally, for the directed polymer
model, which is of special interest in connection with spin glasses in
finite dimensions, we have shown that the droplet picture leads to a
very unusual critical behavior. We have also discussed whether
 the critical temperature could coincide with the upper bound $T_2$ derived by
Derrida and coworkers, depending on the negative tail of the free-energy
distribution in the high temperature phase. 

  \section{ Acknowledgements}
We are grateful to H. Spohn for drawing our attention to Ref. \cite{birkner}
and for interesting comments.
We also thank J. Wehr for correspondence on his work \cite{We_Ai}.

\end{document}